\documentclass[twocolumn]{aastex61}
\usepackage{enumitem}

\DeclareMathAlphabet{\mathitbf}{OML}{cmm}{b}{it}
\newcommand{\Avec}{\mathitbf{A}}

\newcommand{\Apvec}{\mathitbf{A}_{\mathrm{p}}}
\newcommand{\Bvec}{\mathitbf{B}}

\newcommand{\Btrans}{\mathit{B_{\rm hor}}}

\newcommand{\Bpvec}{\mathitbf{B}_{\mathrm{0}}}

\newcommand{\Bobs}{\mathitbf{B}_{\mathrm{obs}}}
\newcommand{\Bpp}{\mathitbf{B}_{\mathrm{pp}}}
\newcommand{\Jvec}{\mathitbf{J}}
\newcommand{\Wvec}{\mathitbf{W}}

\newcommand{\CJT}{$\mathrm{FV}_{\rm Coulomb}$}
\newcommand{\DLL}{$\mathrm{FV}_{\rm DeVore}$}
\newcommand{\Etot}{E}
\newcommand{\Epot}{E_{\mathrm{0}}}
\newcommand{\Efree}{E_{\mathrm{F}}}
\newcommand{\Ediv}{E_{\mathrm{div}}}
\newcommand{\Ejs}{E_{\mathrm{J,s}}}
\newcommand{\Eps}{E_{\mathrm{0,s}}}
\newcommand{\Ej}{E_{\mathrm{J}}}
\newcommand{\Ejns}{E_{\mathrm{J,ns}}}
\newcommand{\Epns}{E_{\mathrm{0,ns}}}
\newcommand{\Emix}{E_{\mathrm{mix}}}
\newcommand{\wa}{w_f}
\newcommand{\wb}{w_d}
\newcommand{\wlos}{w_{\rm los}}
\newcommand{\wtrans}{w_{\rm hor}}
\newcommand{\efprime}{\Efree/\Epot}
\newcommand{\hpj}{H_{\mathrm{PJ}}}
\newcommand{\hj}{H_{\mathrm{J}}}
\newcommand{\hjprime}{|\hj|/|\hv|}
\newcommand{\hv}{H_{\mathcal{V}}}
\newcommand{\runi}{{\sc series~I}}
\newcommand{\runii}{{\sc series~II}}

\newcommand{\thetaj}{\theta_J}
\newcommand{\fiavg}{\langle|f_i|\rangle}

\newcommand{\ie}{{\it i.e.}}
\newcommand{\eg}{{\it e.g.}}

\newcommand{\sdo}{{\it SDO}}
\shortauthors{Thalmann et al.}
\accepted{ApJ Lett.\ on \today}

\begin{document}

\title{On the reliability of magnetic energy and helicity computations based on \\ nonlinear force-free coronal magnetic field models}

\correspondingauthor{Julia K. Thalmann}
\email{julia.thalmann@uni-graz.at}

\author{Julia K. Thalmann}
\affil{University of Graz, Institute of Physics/IGAM, Universit\"atsplatz 5, 8010 Graz, Austria}

\author{L.~Linan}
\author{E.~Pariat}
\affiliation{LESIA, Observatoire de Paris, Universit{\'e} PSL, CNRS, Sorbonne Universit{\'e}, Universit{\'e} de Paris, 5 place Jules Janssen, \\92195 Meudon, France}

\author{G.~Valori}
\affiliation{Mullard Space Science Laboratory, University College London, Holmbury St.\ Mary, Dorking, Surrey RH5 6NT, UK}

\begin{abstract}
We demonstrate the sensitivity of magnetic energy and helicity computations regarding the quality of the underlying coronal magnetic field model. We apply the method of \cite{2010A&A...516A.107W} to a series of {\it SDO}/HMI vector magnetograms, and discuss nonlinear force-free (NLFF) solutions based on two different sets of the free model parameters. The two time series differ from each other concerning their force-free and solenoidal quality. Both force- and divergence-freeness are required for a consistent NLFF solution. Full satisfaction of the solenoidal property is inherent in the definition of relative magnetic helicity in order to insure gauge-independence. We apply two different magnetic helicity computation methods \citep[][]{2011SoPh..272..243T, 2012SoPh..278..347V} to both NLFF time series and find that the output is highly dependent on the level to which the NLFF magnetic fields satisfy the divergence-free condition, with the computed magnetic energy being less sensitive than the relative helicity. Proxies for the non-potentiality and eruptivity derived from both quantities are also shown to depend strongly on the solenoidal property of the NLFF fields. As a reference for future applications, we provide quantitative thresholds for the force- and divergence-freeness, for the assurance of reliable computation of magnetic energy and helicity, and of their related eruptivity proxies.
\end{abstract}

\keywords{Sun: corona -- Sun: flares -- Sun: magnetic fields -- methods: data analysis -- methods: numerical}

\section{Introduction}\label{sec:introduction} 

For practical cases, \cite{2012SoPh..278..347V} demonstrated the validity and physical meaningfulness to compute (and track in time) the relative magnetic helicity in order to characterize (the evolution of) a magnetic system. As its name implies, the relative helicity allows it to express the helicity of a magnetic field with respect to a reference field. This relative formulation allows it to circumvent the problem that magnetic helicity cannot be defined meaningfully for systems that are not magnetically closed (such as the solar corona). 

Following \cite{1984JFM...147..133B} and \cite{1984CPPCF...9..111F}, the relative magnetic helicity (simply called helicity hereafter) in a volume, $\mathcal V$, bounded by a surface, $\partial\mathcal V$, can be written as 
\begin{equation}
	\hv=\int_\mathcal{V}\left(\Avec+\Apvec\right)\cdot\left(\Bvec-\Bpvec\right) {\rm ~d}\mathcal{V}, \label{eq:hv}
\end{equation}
where the reference field, $\Bpvec$, shares the normal component of the studied field $\Bvec$ on $\partial\mathcal{V}$. Often, a potential (current-free) field is used as reference field \citep[see][for an alternative choice]{2014ApJ...787..100P}. In \href{eq:hv}{Eq.~(\ref{eq:hv})}, $\Avec$ and $\Apvec$ are the vector potentials satisfying $\Bvec=\nabla\times\Avec$ and $\Bpvec=\nabla\times\Apvec$, respectively.

Following
 \cite{1999PPCF...41B.167B}, \href{eq:hv}{Eq.~(\ref{eq:hv})} may be written as $\hv=\hj+\hpj$, with
\begin{eqnarray}
	\hj&=&\int_\mathcal{V}\left(\Avec-\Apvec\right)\cdot\left(\Bvec-\Bpvec\right) {\rm ~d}\mathcal{V}, \label{eq:hj}\\
	\hpj&=&2\int_\mathcal{V}\Apvec\cdot\left(\Bvec-\Bpvec\right) {\rm ~d}\mathcal{V}. \label{eq:hpj}
\end{eqnarray}
Here, $\hj$ is the magnetic helicity in the volume associated to the electric current, and $\hpj$ is the helicity associated with the component of the field that is threading $\partial\mathcal V$. Because $\Bvec$ and $\Bpvec$ are designed such that they share their normal component, $\Bvec_n$, on $\partial\mathcal{V}$, not only $\hv$ but also both, $\hj$ and $\hpj$ are independently gauge invariant. 

Importantly, the underlying magnetic fields, $\Bvec$ and $\Bpvec$, have to adhere to a certain level of divergence freeness, in order to ensure reliable helicity computation. For this purpose, \cite{2013A&A...553A..38V} used the decomposition of the magnetic energy within $\mathcal{V}$ in the form
\begin{eqnarray}
\Etot&=&\frac{1}{2\mu_0}\int_\mathcal{V} B^2 {\rm ~d}\mathcal{V} = \Epot + \Ej \nonumber \\
&=&\Eps+\Ejs+\Epns+\Ejns+\Emix, \label{eq:e_i}
\end{eqnarray}
with $\Epot$ and $\Ej$ being the energies of the potential and current-carrying magnetic field, respectively. $\Epot$ is used to compute an upper limit for the free energy as $\Efree=\Etot-\Epot$. $\Eps$ and $\Ejs$ are the energies of the potential and current-carrying solenoidal magnetic field components. $\Epns$ and $\Ejns$ are those of the corresponding non-solenoidal components. $\Emix$ corresponds to all cross terms \citep[see Eq.~(8) in][for the detailed expressions]{2013A&A...553A..38V}. For a perfectly solenoidal field, one finds $\Eps=\Epot$, $\Ejs=\Ej$, and $\Epns=\Ejns=\Emix=0$.

Based on \href{eq:e_i}{Eq.~(\ref{eq:e_i})}, \cite{2016SSRv..201..147V} introduced the ratio $\Ediv/\Etot$, with $\Ediv=\Epns+\Ejns+|\Emix|$, as to be indicative of the divergence-freeness of the magnetic field, and tested the corresponding sensitivity of \href{eq:hv}{Eq.~(\ref{eq:hv})}, based on different numerical methods to compute magnetic helicity. Based on a specifically designed numerical experiment, where a finite divergence was added in a controlled way to a numerically solenoidal MHD model case, it was shown that the error in the computation of $\hv$ may grow considerably, if $\Ediv/\Etot\gtrsim0.1$.

Magnetic helicity computations are often performed based on nonlinear force-free (NLFF) coronal magnetic field extrapolations, using the optimization method of \cite{2010A&A...516A.107W}. It represents the numerical solution to the boundary value problem of extrapolation of the measured surface magnetic field into the solar corona,
\begin{eqnarray}
	(\nabla\times\Bvec)\times\Bvec &=& \mathbf{0} \label{eq:ff1}\\
	\nabla\cdot\Bvec&=&0.\label{eq:ff2}
\end{eqnarray}

The method allows several free model parameters to be chosen, in order to optimize the numerical solution of \href{eq:ff1}{Eqs.~(\ref{eq:ff1})} and (\ref{eq:ff2}). If used as a black box, pre-defined values are used, without optimization regarding underlying specific magnetogram data. The pre-defined values are to be thought of as to be initial guesses only, however, and a careful testing and tuning of the free model parameters is inevitable. Only a careful selection of the free model parameters is capable of producing NLFF solutions of highest quality, both, in terms of force- and divergence-freeness. A high degree of force-freeness is crucial for the validity of the NLFF solution with respect to the measured photospheric field it is based on. A low level of divergence is mandatory for a reliable computation of $\hv$. 

The presented work represents an extension of the work by \cite{2016SSRv..201..147V}, by considering the dependency of energy and helicity computations on the field's solenoidal property in observed solar cases, rather than an idealized model. Our work shall serve as a reference, concerning the quality a NLFF model has to suffice, in order to be used as an input for reliable helicity modeling.

\section{Method}

We use photospheric vector magnetic field data \citep{2014SoPh..289.3483H}, derived from {\it Solar Dynamics Observatory} \citep[\sdo;][]{2012SoPh..275....3P} Helioseismic and Magnetic Imager \citep[HMI;][]{2012SoPh..275..229S} polarization measurements. We use the {\sc hmi.sharp\_cea\_720s} data series which provides a Lambert Cylindrical Equal-Area projected magnetic field vector within automatically-identified active region patches \citep[][]{2014SoPh..289.3549B}, with the azimuthal component of the vector magnetic field being disambiguated \citep[][]{1994SoPh..155..235M,2009SoPh..260...83L}.

For computational feasibility, we bin the photospheric data by a factor of 4 to a plate scale of $0.12$~degree. We set our analysis time range such that it covers the time period 2011 February~12 to 16, \ie, the disk passage of active region (AR) NOAA~11158. Around intense flares (equal or larger {\it GOES} class M5.0), we use HMI's native time cadence of 12 minutes, and an 1-hour cadence otherwise. 

Based on the binned vector magnetic field data, we compute NLFF equilibria for each time step, which involves two computational steps. Firstly, we ``preprocess'' the data, to obtain a more force-free consistent state \citep{2006SoPh..233..215W}. The preprocessing method allows different free parameters to be set:
\begin{itemize}[itemsep=1pt]
\item[--] $\mu_1$ and $\mu_2$ control the level of force and torque of the data,
\item[--] $\mu_3$ allows deviations from the input data, and 
\item[--] $\mu_4$ controls the degree of applied smoothing.
\end{itemize} 
In the original notation the pre-defined standard setting is ($\mu_1$,$\mu_2$,$\mu_3$,$\mu_4$)=($1$,$1$,$10^{-3}$,$10^{-2}$). 

Secondly, we apply the method of \cite{2010A&A...516A.107W} to the preprocessed maps. The optimization approach is designed such that the functional
\begin{eqnarray}
	L &=& \int_V \wa\,\frac{|\left(\nabla\times\Bvec\right)\times\Bvec|^2}{B^2}+\wb\,|\nabla\cdot\Bvec|^2\,{\rm d}v  \nonumber \\
	&+& \nu \int_S(\Bvec-\Bobs)\cdot\Wvec\cdot\left(\Bvec-\Bobs\right)\,{\rm d}s, \label{eq:opt}
\end{eqnarray}
is minimized such that the volume-integrated Lorentz force and divergence becomes small. 

The surface term in \href{eq:opt}{Eq.~(\ref{eq:opt})} allows deviations between the NLFF solution, $\Bvec$, and the magnetic field information at the lower boundary, $\Bobs$, in order to find a more force-free solution. The deviation from $\Bobs$ is controlled by the diagonal error matrix, $\Wvec$, which allows it to incorporate uncertainties on each component of the magnetic field, and in each pixel, separately. Ideally, $\Bobs$ would be a magnetogram measured at a chromospheric height, \ie, in a force-free regime of the solar atmosphere. In practice, the preprocessed photospheric vector field, $\Bpp$ is supplied, so that $\Bobs=\Bpp$.

The model parameters that can be freely assigned in \href{eq:opt}{Eq.~(\ref{eq:opt})} are:
\begin{itemize}[itemsep=1pt]
\item[--] Separate weightings of the volume-integrated force ($\wa$) and divergence ($\wb$). In the original notation these are set as $\wa=\wb=1$.
\item[--] The components $\wlos$ and $\wtrans$ of the diagonal error matrix, $\Wvec$, can be defined in different ways. The choice $\wlos=\wtrans=1$ assures accuracy of both, the longitudinal and horizontal magnetic field, equally at all pixel locations. Alternatively, $\wtrans=|\Btrans| / {\rm max}(|\Btrans|)$ may be applied, \ie, assuming stronger horizontal fields to be measured with higher accuracy.  \item[--] The impact of the surface term, \ie, the influence of $\Bobs$ onto the final NLFF solution, is controlled by $\nu$. \cite{2010A&A...516A.107W}, suggest $\nu$ in the range $10^{-4}$--$10^{-1}$.
\end{itemize}

Successful NLFF modeling involves to find a combination of free model parameters that delivers optimized results, in terms of force- and divergence-freeness. In order to quantify the consistency of the obtained NLFF solutions, we use the current-weighted average of the angle between the modeled magnetic field and electric current density, $\thetaj$, \citep[][]{2006SoPh..235..161S}. We use the volume-averaged fractional flux, $\fiavg$, \citep[][]{2000ApJ...540.1150W} and the energy ratio, $\Ediv/E$, \citep{2013A&A...553A..38V}, to quantify the level of divergence of the NLFF solution.

We use two finite-volume (FV) methods to compute the magnetic helicity based on \href{eq:hv}{Eqs.~(\ref{eq:hv})}--\href{eq:hv}{(\ref{eq:hpj})}. The method of \cite{2011SoPh..272..243T}, solves systems of partial differential equations to obtain the vector potentials $\Avec$ and $\Apvec$, using the Coulomb gauge, $\nabla\cdot\Avec=\nabla\cdot\Apvec=0$, (``\CJT'', hereafter). The method of \cite{2012SoPh..278..347V} is based on integral formulations, using the DeVore gauge, $A_z=A_{{\rm p},z}=0$, (``\DLL'', hereafter). { Both methods define the reference field as $\Bpvec=\nabla\phi$, with $\phi$ being the scalar potential, subject to the constraint $\nabla_n\phi=\Bvec_n$ on $\partial\mathcal{V}$.} The methods have been tested in the framework of an extended proof-of-concept study on FV helicity computation methods \citep[][]{2016SSRv..201..147V}, where it has been shown that for various test setups the methods deliver helicity values in line with each other, differing by a few percent only.

In total we employed 18 NLFF extrapolations of the same HMI magnetogram at 12:00~UT, obtained with 18 different combinations of the parameters ($\mu_3$,$\mu_4$,$\wb$,$\wtrans$), in order to pin down successful parameter sets. Unphysical NLFF solutions with $\Epot<\Etot$, and solutions where $\Ediv/\Etot>0.1$ were discarded from further consideration. Seven solutions were found with favorable properties for helicity modeling, with $\Ediv/\Etot\lesssim0.1$. Out of those, two parameters sets were chosen for further consideration. In combination with ($\mu_1$, $\mu_2$, $\wa$, $\nu$)=(1, 1, $1$, $10^{-3}$), our first selected choice ($\mu_3$, $\mu_4$, $\wb$, $\wtrans$)= ($10^{-3}$, $10^{-3}$, 2, $\propto\Btrans$) delivered a NLFF solution with an exceptionally low solenoidal level ($\Ediv/\Etot<0.01$). A NLFF model solution close to the limit $\Ediv/\Etot\simeq0.1$, suggested as to be tolerable for helicity modeling in \cite{2016SSRv..201..147V}, was found based on the choice ($\mu_3$, $\mu_4$, $\wb$, $\wtrans$)=($10^{-3}$, $10^{-3}$, 1, 1), and also selected for further analysis.

We used these two sets of free model parameters to compute the full time series of NLFF models for NOAA~11158 between 12~February 00:00~UT and 16~February 00:00~UT (hereafter called \runii\ and \runi\, respectively), in order to demonstrate how important the degree of $\nabla\cdot\Bvec$ of the input NLFF solutions is for successful helicity computation.

\begin{figure*}[t]
	\centering
	\includegraphics[width=\textwidth]{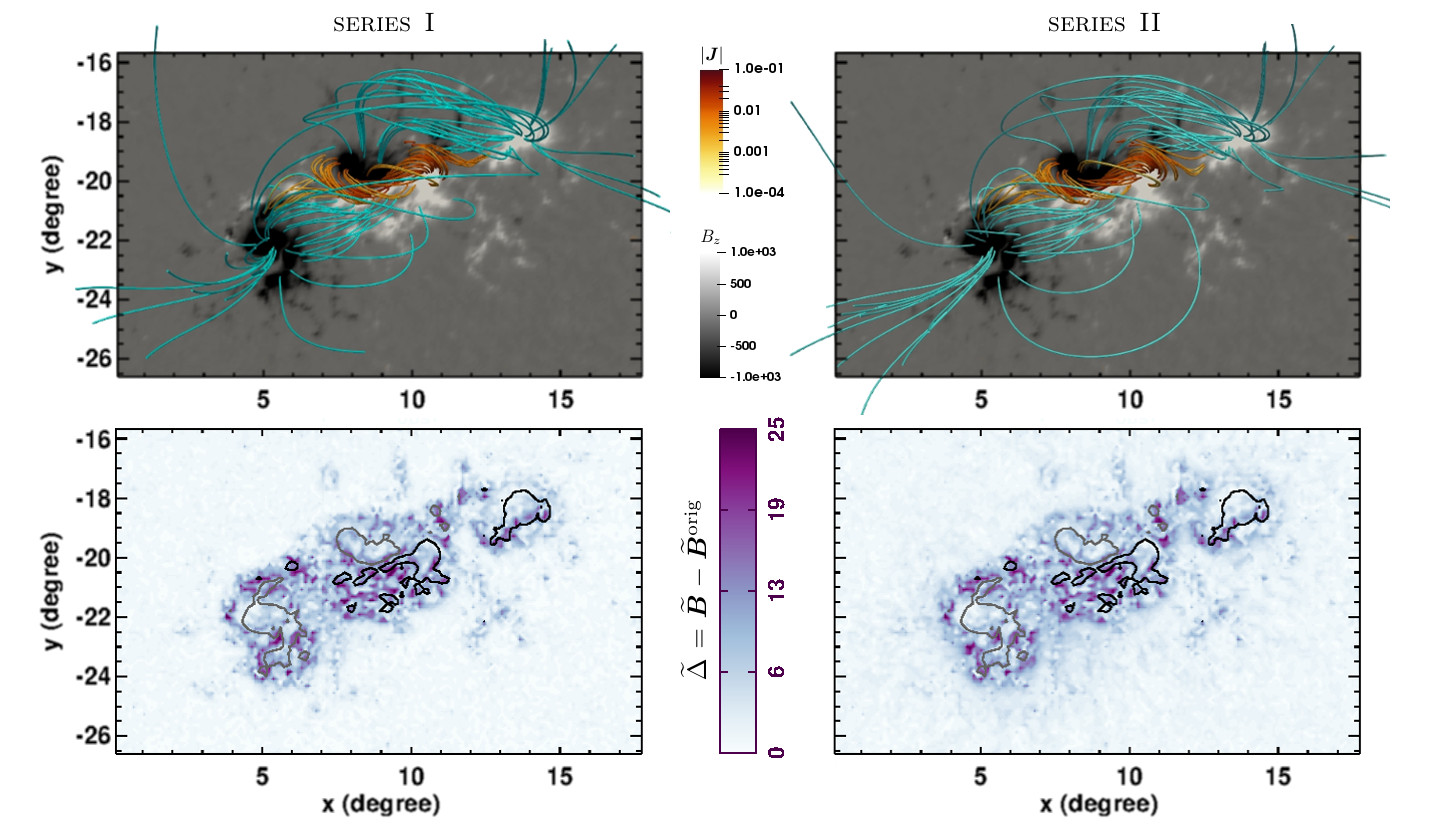}
	\put(-465,262){\bf\color{white}(a)}
	\put(-208,262){\bf\color{white}(b)}
	\put(-465,125){\bf(c)}
	\put(-208,125){\bf(d)}
\caption{
NLFF model solution for February~14 at 21:00~UT in (a) \runi\ and (b) \runii. Sample field lines outlining the large-scale magnetic field are colored green. Those originating from the AR center are color-coded according to the total absolute current density, $|\Jvec|$, at their footpoints. The gray scale background shows the measured vertical magnetic field, $\mathit{B}_z$, scaled to $\pm\,1$~kG. Panels (c) and (d) show the respective normalized changes at the lower NLFF boundary. Black/gray contours are drawn at $\pm750$~G.\\
}
\label{fig:fig1}
\end{figure*}

\begin{figure*}[t]
	\centering
	\includegraphics[width=\textwidth]{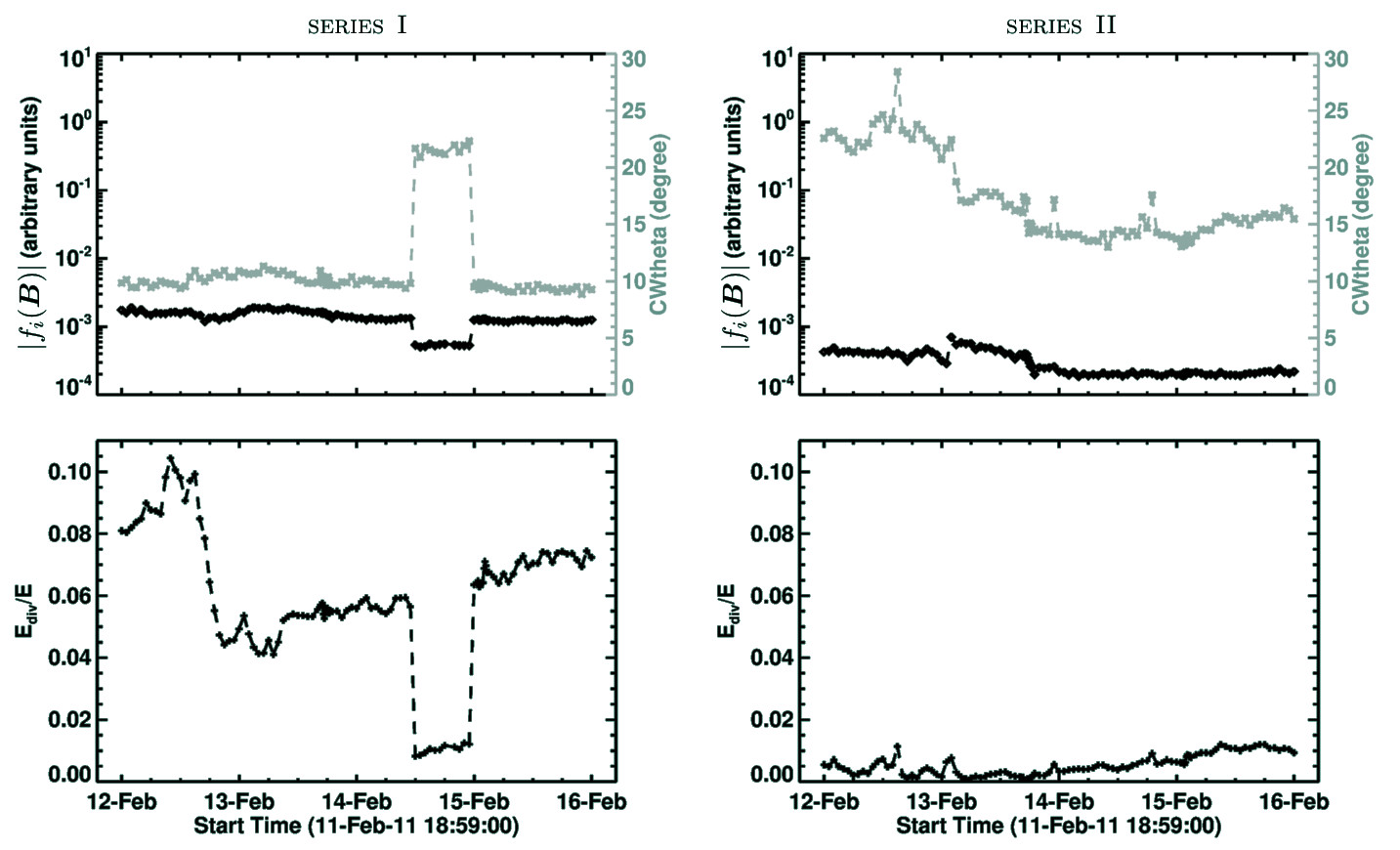}
	\put(-470,278){\bf(a)}
	\put(-213,278){\bf(b)}
	\put(-470,135){\bf(c)}
	\put(-213,135){\bf(d)}
\caption{
Quality of NLFF magnetic fields of \runi\ (left panels) and \runii\ (right panels). $\thetaj$ (gray stars) in panels (a) and (b) quantifies the degree of force-freeness. The fractional flux, $\fiavg$ (black triangles) quantifies the level of $\nabla\cdot\Bvec$. The non-solenoidal contribution, $\Ediv$, to the total energy is shown in panels (c) and (d).\\ 
}
\label{fig:fig2}
\end{figure*}

\section{Results}
\label{sec:results}

\subsection{Properties and quality of NLFF modeling}

In \href{fig:fig1}{Fig.~\ref{fig:fig1}a} and \href{fig:fig1}{\ref{fig:fig1}b}, the NLFF model solution of \runi\ and \runii\ for February~14 21:00~UT are shown, respectively. Both reveal a low-lying system of helical magnetic field along the main polarity inversion line in the AR center, in agreement with earlier works \citep[\eg,][]{2012ApJ...752L...9J,2012ApJ...748...77S,2013ApJ...770...79I}.

Based on the used free parameter sets, the binned vector magnetic field, $\Bvec^{\rm orig}$, is changed to a different degree during NLFF modeling. Following, \cite{2015ApJ...811..107D}, we characterize the modifications of the vector field at the lower boundary ($z=0$) as
\begin{equation}
\Delta=\Bvec_i-\Bvec_i^{\rm orig},
\end{equation}
for each component $i=\{x,y,z\}$, where $\Bvec$ is the final magnetic field at the lower boundary of the NLFF model. The magnitudes of the changes are considered separately for the vertical ($\Delta_z$) and horizontal magnetic field $\Delta_h=|(\Delta_x,\Delta_y)|$. In addition, we use the normalized change, $\widetilde\Delta=\widetilde{\Bvec}-\widetilde{\Bvec}^{\rm orig}$, to incorporate the measurement uncertainties for each component, $\sigma_i$, where $\widetilde{\mathit{B}}_i=\mathit{B}_i/\sigma_i$ and 
$\widetilde{\mathit{B}}_i^{\rm orig}=\mathit{B}^{\rm orig}_i/\sigma_i$, 
for the computation of $\widetilde\Delta_z$ and $\widetilde\Delta_h=|(\widetilde\Delta_x,\widetilde\Delta_y)|/\sqrt{2}$. \href{tab:nlff_changes}{Table~\ref{tab:nlff_changes}} lists the rms values for the magnitudes of the (normalized) changes of the vertical and horizontal magnetic field components. The normalized changes tend to be larger in weak field regions, \ie, outside of the AR core (see \href{fig:fig1}{Fig.~\ref{fig:fig1}c}, \href{fig:fig1}{\ref{fig:fig1}d}).

\begin{deluxetable}{ccccc}
\tablecaption{
Changes to the measured magnetic field during NLFF modeling, as shown in \href{fig:fig1}{Fig.~\ref{fig:fig1}c} and \href{fig:fig1}{\ref{fig:fig1}d}.
\label{tab:nlff_changes}
}
\tablehead{
 \multicolumn{1}{c}{Series} & $\Delta_z^{rms}$ & $\Delta_h^{rms}$ & $\widetilde\Delta_z^{rms}$ & $\widetilde\Delta_h^{rms}$\\
~ & (G) & (G) & (G) & (G)
}
\startdata
I & 68.74 & \color{white}1\color{black}81.36 & 4.30 & 2.42 \\
II & 63.03 & 103.54 & 4.06 & 2.88 \\
\enddata
\end{deluxetable}

\href{fig:fig2}{Fig.~\ref{fig:fig2}a} shows the mean current-weighted angle for \runi\ (gray stars), with a median of $\thetaj=9.8^\circ\pm1.5^\circ$. For the volume-averaged fractional flux (black triangles), we find a median value of $\fiavg\times10^4=13.3\pm2.5$. \href{fig:fig2}{Fig.~\ref{fig:fig2}c} shows that $\Ediv/\Etot\gtrsim0.05$, with a median value of $\Ediv/\Etot=0.06\pm0.02$, for the majority of time instances considered. For \runii, we find the median values $\thetaj=15.6^\circ\pm2.7^\circ$ and $\fiavg\times10^4=2.2\pm1.0$. The non-solenoidal contribution to the total energy is considerably lower than in \runi\ (\href{fig:fig2}{Fig.~\ref{fig:fig2}d}), with a median value $\Ediv/\Etot=0.005\pm0.003$. Note that the improved solenoidal condition in \runii\ is achieved on the slight expense of force-freeness.

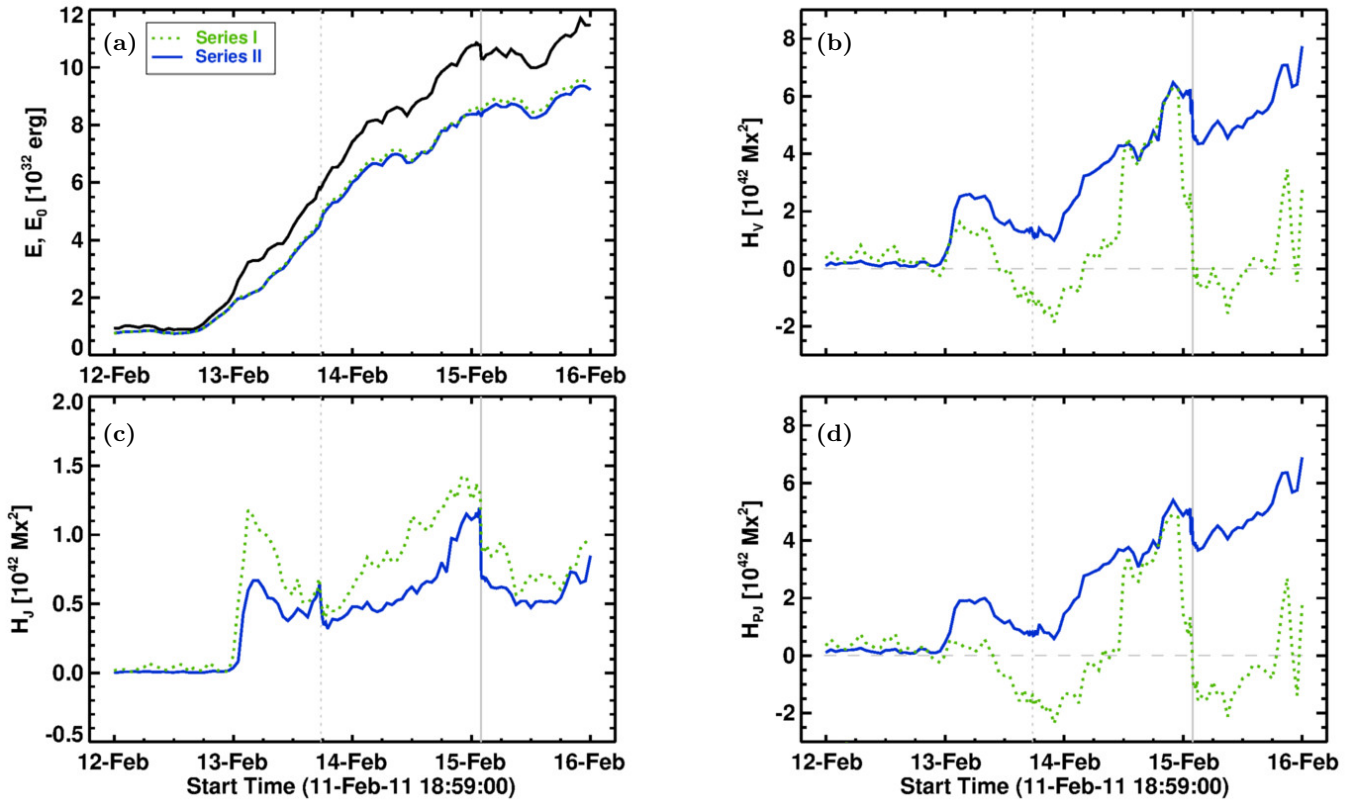
\begin{figure*}
	\centering
	\includegraphics[width=\textwidth]{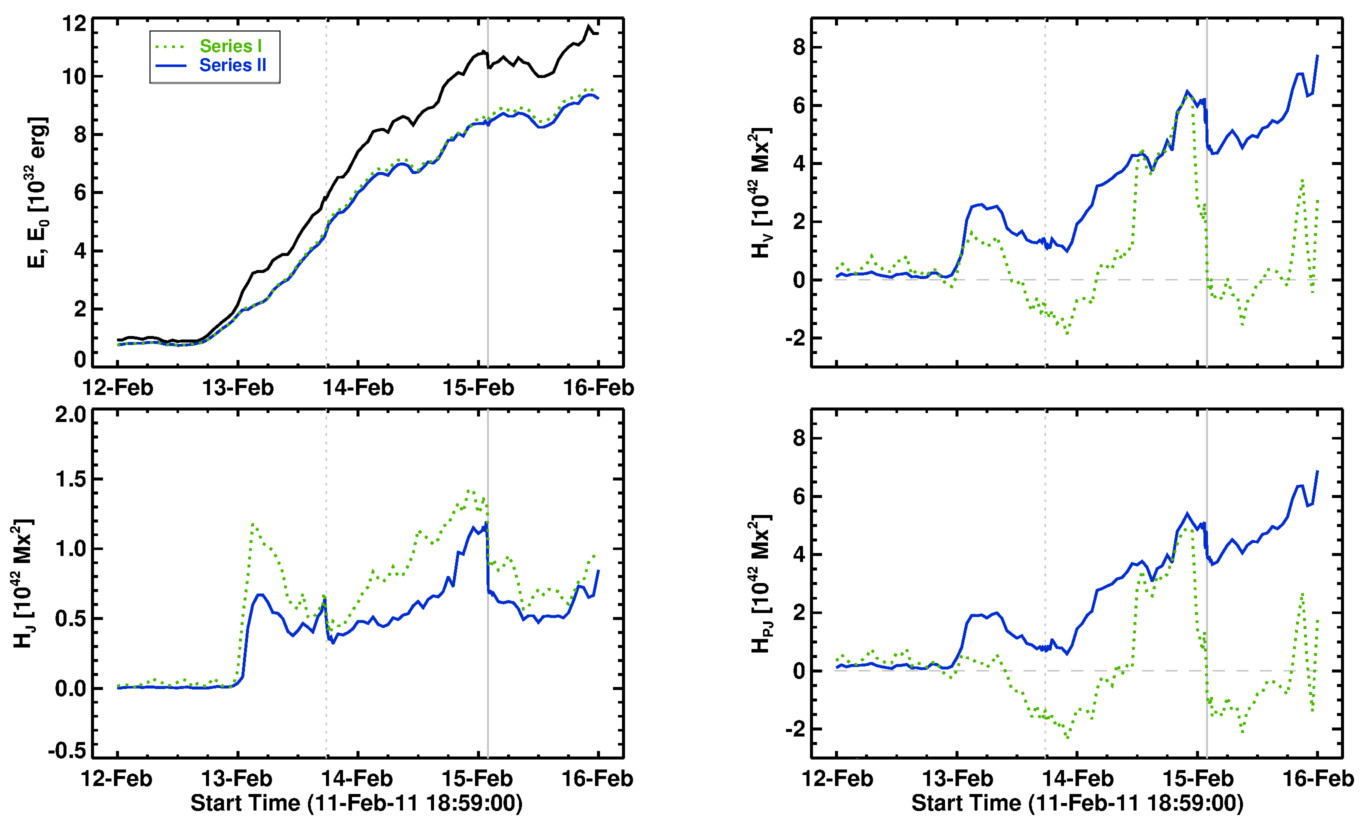}
	\put(-474,288){\bf(a)}
	\put(-204,288){\bf(b)}
	\put(-474,140){\bf(c)}
	\put(-204,140){\bf(d)}
\caption{
(a) Total energy ($\Etot$; black solid line) and potential energy ($\Epot$) for \runi\ (green dashed) and \runii\ (blue solid). Panel (b) shows the corresponding total helicity, $\hv$, derived using the \CJT\ method. In panels (c) and (d), the contributions of $\hj$ and $\hpj$ are shown, respectively. Vertical dashed and solid lines mark the {\it GOES} peak time of M- and X-class flares, respectively.\\
}
\label{fig:fig3}
\end{figure*}

\subsection{Effect of divergence on helicity computations}

Fast-evolving NOAA~11158 showed a considerable increase of unsigned magnetic flux starting on late February~12, at a time when a pronounced filament started to emerge \citep[for an in-depth analysis see][]{2012ApJ...748...77S}. Parts of the filament erupted during two eruptive flares, an M6.6 flare (SOL2011-02-13T17:38) and an X2.2 flare (SOL2011-02-15T01:56). \href{fig:fig3}{Fig.~\ref{fig:fig3}a} shows the corresponding total ($\Etot$; black solid line) and potential field ($\Epot$) energy for \runi\ (green dotted line) and \runii\ (blue solid line), including a considerable increase, starting with the emergence of the filament early on February~13.

\href{fig:fig3}{Fig.~\ref{fig:fig3}b} shows the total helicity, $\hv$, computed with the \CJT\ method. While exclusively positive values are found for $\hv$ when based on \runii, the corresponding curve of \runi\ shows an unexpected behavior, including rapid and drastic changes, independent of the occurrence of the eruptive flares. A closer look into the contributors to the total helicity, $\hj$ (\href{fig:fig3}{Fig.~\ref{fig:fig3}c}) and $\hpj$ (\href{fig:fig3}{Fig.~\ref{fig:fig3}d}) reveals that these changes mainly stem from the contribution of $\hpj$. The magnitudes of $\hj$ differ less, with slightly lower values obtained from \runii. 

Though not shown explicitly, we note that the results derived using the \DLL\ method show a similar behavior, though slightly less extreme. In \runi, a good match of $\hv$ (and thus $\hpj$) between the two methods is only found during a short time interval, between February~14 $\sim$12:00~UT and early February~15, \ie~where the input magnetic fields were more divergence free ($\fiavg\propto10^{-4}$ and $\Ediv/\Etot\simeq0.01$) than at other times (compare \href{fig:fig2}{Fig.~\ref{fig:fig2}c}). In contrast, \CJT\ and \DLL\ deliver almost identical results for the entire \runii.

\begin{figure*}
	\centering
	\includegraphics[width=\textwidth]{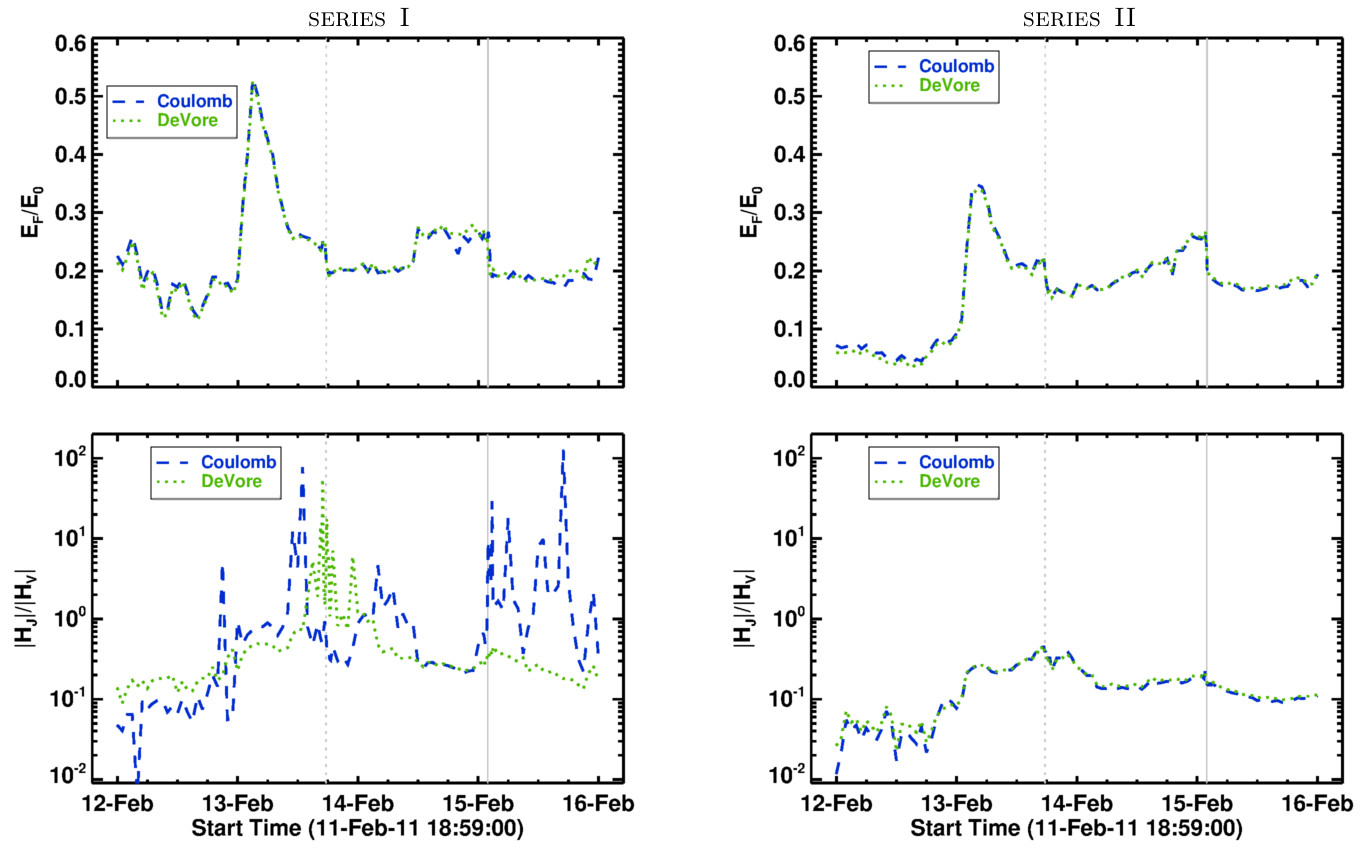}
	\put(-474,289){\bf(a)}
	\put(-204,289){\bf(b)}
	\put(-474,140){\bf(c)}
	\put(-204,140){\bf(d)}
\caption{
Magnetic energy ratio, $\efprime$, for (a) \runi\ and (b) \runii. Panels (c) and (d) show the helicity ratio, $\hjprime$, respectively. Blue dashed and green dotted lines represent the model solutions based on the \CJT\ and \DLL\ method. Vertical dashed and solid lines mark the {\it GOES} peak time of M- and X-class flares, respectively.\\
}
\label{fig:fig4}
\end{figure*}

\subsection{Effect of divergence onto eruptivity proxies}

Often employed are proxies quantifying the non-potentiality and eruptivity in the form of the free energy ratio, $\efprime$, and the helicity ratio, $\hjprime$. In \href{fig:fig4}{Fig.~\ref{fig:fig4}}, we compare the effect of divergence in \runi\ (left panels) and \runii\ (right panels) onto the corresponding values derived with the \CJT\ and \DLL\ method.

While $\efprime$ is growing prior to the eruptive flares and $\gtrsim0.2$ in \runii\ (\href{fig:fig4}{Fig.~\ref{fig:fig4}b}), this is not the case for \runi\ (\href{fig:fig4}{Fig.~\ref{fig:fig4}a}). Also, in \runi, $\efprime$ appears rather large prior to the presence of strong magnetic fluxes (before late February~12) at a time when the AR did not exhibit signatures of a filament. Thus, the energy (ratio) at those times is mostly dominated by the level of $\nabla\cdot\Bvec$, and in that sense, indicates that anything below $\efprime\simeq0.2$ is not significant.

The effect on the helicity ratio $\hjprime$ is equally severe. While it may serve as a proxy for eruptivity, with (in-) decreasing trend (before) after the major flares in \runii, and values $\hjprime\gtrsim0.2$ prior to flare occurrence (\href{fig:fig4}{Fig.~\ref{fig:fig4}d}), such a conclusion cannot be drawn based on \runi\ (\href{fig:fig4}{Fig.~\ref{fig:fig4}c}). Here, clear and smooth trends around flares are hard to discriminate. In particular, $\hjprime$ exceeds a value of one at times when $\hpj$ based on the respective method takes unexpected turns (compare the dashed blue line in \href{fig:fig4}{Fig.~\ref{fig:fig4}c} and the green dotted line in \href{fig:fig3}{Fig.~\ref{fig:fig3}e} for the \CJT\ method). Again, $\hjprime$ based on the two different methods agrees only for times in \runi\ when the underlying magnetic field is more divergence free (compare \href{fig:fig2}{Fig.~\ref{fig:fig2}a} and \href{fig:fig2}{\ref{fig:fig2}c}).

\section{Summary and Discussion}

We aimed at testing the sensitivity of magnetic energy and helicity computations, based on real solar observations, in terms of the quality of the employed coronal magnetic field model. We employed the method of \cite{2010A&A...516A.107W} based on two different free model parameter sets to a time series of observed vector magnetograms, in order to obtain two time series of NLFF models (\runi\ and \runii). A high degree of force-freeness is crucial for the validity of a NLFF solution with respect to the measured photospheric field it is based on. A low level of divergence is mandatory for a reliable computation of magnetic helicity \citep{2013A&A...553A..38V,2016SSRv..201..147V}.

While the NLFF fields of \runi\ are of ``standard'' quality ($\thetaj\simeq10^\circ$, $\fiavg\propto10^{-3}$, and $\Ediv/\Etot\gtrsim0.05$), those are of \runii\ are of higher solenoidal quality ($\fiavg\propto10^{-4}$ and $\Ediv/\Etot\simeq0.01$) and slightly lower force-freeness (\href{fig:fig2}{Fig.~\ref{fig:fig2}}).
The numbers for \runii\ represent remarkably good values for observation-based NLFF modeling, competing with the best-performing models discussed in \cite{2015ApJ...811..107D}, while better preserving the original input vector magnetic field \citep[compare the rms values in our \href{tab:nlff_changes}{Table~\ref{tab:nlff_changes}} and Table~3 in][]{2015ApJ...811..107D}. 

We applied two different FV methods to compute the magnetic helicity \citeauthor{2011SoPh..272..243T}\,(\citeyear{2011SoPh..272..243T}; ``\CJT'') and \citeauthor{2012SoPh..278..347V}\,(\citeyear{2012SoPh..278..347V}; ``\DLL'') which are based on different gauges (Coulomb vs.\ DeVore, respectively) and employ different mathematical approaches (differential vs.\ integral formulation for the vector potentials, respectively). We applied both methods to both NLFF time series and find that the different methods deliver almost identical results (\ie, $\hv$, as well as $\hj$ and $\hpj$), given a sufficient solenoidal quality of the input magnetic field.  We therefore suggest that, quite generally, helicity computations may be meaningful and trustworthy only, if $\Ediv/\Etot\lesssim0.05$ and $\fiavg\times10^4\lesssim5$, for the underlying magnetic field model.

The different methods react differently on the quality of the input fields, with \CJT\ being more sensitive, with larger absolute variations in $\hpj$ (\href{fig:fig3}{Fig.~\ref{fig:fig3}d}), and hence $\hv$ (\href{fig:fig3}{Fig.~\ref{fig:fig3}b}). The least difference, even for non-negligible divergence, is found for $\hj$, both, between the methods and between the two NLFF series (\href{fig:fig3}{Fig.~\ref{fig:fig3}c}), which seems to point to an inconsistency in context with the potential field. In comparison, the magnetic energy shows only little sensitivity to the quality of the underlying NLFF solution (\href{fig:fig3}{Fig.~\ref{fig:fig3}a}). Irrespective of the method, the proxies for non-potentiality, $\efprime$, and for eruptivity, $\hjprime$, are affected to a degree which allows reliable conclusions only if the input NLFF field is solenoidal enough (\href{fig:fig4}{Fig.~\ref{fig:fig4}}).

In our case, the unexpected behavior of $\hpj$ (and thus, $\hv$) in \runi\ is caused by a too large divergence of the underlying NLFF solutions ($\fiavg\times10^4\gtrsim5$ and $\Ediv/\Etot\gtrsim0.05$). Correspondingly, we are able to verify the doubts of \cite{2014SoPh..289.4453M} concerning the reliability of their long-term helicity analysis of ARs NOAA~11072 ($\fiavg\times10^3=1.3\pm0.2$) and NOAA~11158 ($\fiavg\times10^4=7.2\pm0.9$). Corresponding judgment of other earlier works are difficult, because relevant control parameters were not reported \citep[\eg,][]{2012ApJ...752L...9J,2015RAA....15.1537J}.

The effect of non-solenoidal contributions to $\hv$ may be case-dependent, however. A similar behavior may in some cases just represent the correct evolution. For instance, the most eruptive case of MHD simulations analyzed in \cite{2017A&A...601A.125P} shows a change of the sign of $\hpj$, though smoothly and to values small compared to the pre-eruption value. Also, a variation of the sign of $\hj$ around zero, around times when strong magnetic flux is initially emerging in an AR, may just be physical \citep[compare our \href{fig:fig3}{Fig.~\ref{fig:fig3}} and Fig.~6 of][]{2017A&A...601A.125P}.

In summary, we find that a quantitative assessment of the consistency of NLFF models in terms of force- and divergence-freeness is mandatory for making any reliable statement involving their energy and helicity content. Moreover, despite the necessity of high-quality (\ie, low-divergence) input magnetic fields for helicity computation, simplistic interpretations of the computed magnetic helicity of complex magnetic systems should be taken with care. 

{\footnotesize ~\\
We thank the anonymous referee for helpful comments. J.\,K.\,T.\ acknowledges Austrian Science Fund (FWF): P31413-N27. E.\,P.\ and L.\,L.\ acknowledge support of the French Agence Nationale pour la Recherche through the HELISOL project ANR-15-CE31-0001. G.\,V.\ acknowledges the Leverhulme Trust Research Project Grant 2014-051. {\it SDO} data are courtesy of the NASA/{\it SDO} AIA and HMI science teams. This article profited from discussions during the ISSI International Team meetings on {\it Magnetic Helicity estimations in models and observations of the solar magnetic field} and {\it Magnetic Helicity in Astrophysical Plasmas}.
}


\begin{thebibliography}{}
\expandafter\ifx\csname natexlab\endcsname\relax\def\natexlab#1{#1}\fi
\providecommand{\url}[1]{\href{#1}{#1}}

\bibitem[{{Berger}(1999)}]{1999PPCF...41B.167B}
{Berger}, M.~A. 1999, Plasma Physics and Controlled Fusion, 41, B167

\bibitem[{{Berger} \& {Field}(1984)}]{1984JFM...147..133B}
{Berger}, M.~A., \& {Field}, G.~B. 1984, Journal of Fluid Mechanics, 147, 133

\bibitem[{{Bobra} {et~al.}(2014){Bobra}, {Sun}, {Hoeksema}, {Turmon}, {Liu},
  {Hayashi}, {Barnes}, \& {Leka}}]{2014SoPh..289.3549B}
{Bobra}, M.~G., {Sun}, X., {Hoeksema}, J.~T., {et~al.} 2014, \solphys, 289,
  3549

\bibitem[{{DeRosa} {et~al.}(2015){DeRosa}, {Wheatland}, {Leka}, {Barnes},
  {Amari}, {Canou}, {Gilchrist}, {Thalmann}, {Valori}, {Wiegelmann},
  {Schrijver}, {Malanushenko}, {Sun}, \& {R{\'e}gnier}}]{2015ApJ...811..107D}
{DeRosa}, M.~L., {Wheatland}, M.~S., {Leka}, K.~D., {et~al.} 2015, \apj, 811,
  107

\bibitem[{{Finn} \& {Antonsen}(1984)}]{1984CPPCF...9..111F}
{Finn}, J., \& {Antonsen}, T.~J. 1984, Comments Plasma Phys. Controlled Fusion,
  9, 111

\bibitem[{{Hoeksema} {et~al.}(2014){Hoeksema}, {Liu}, {Hayashi}, {Sun},
  {Schou}, {Couvidat}, {Norton}, {Bobra}, {Centeno}, {Leka}, {Barnes}, \&
  {Turmon}}]{2014SoPh..289.3483H}
{Hoeksema}, J.~T., {Liu}, Y., {Hayashi}, K., {et~al.} 2014, \solphys, 289, 3483

\bibitem[{{Inoue} {et~al.}(2013){Inoue}, {Hayashi}, {Shiota}, {Magara}, \&
  {Choe}}]{2013ApJ...770...79I}
{Inoue}, S., {Hayashi}, K., {Shiota}, D., {Magara}, T., \& {Choe}, G.~S. 2013,
  \apj, 770, 79

\bibitem[{{Jing} {et~al.}(2012){Jing}, {Park}, {Liu}, {Lee}, {Wiegelmann},
  {Xu}, {Deng}, \& {Wang}}]{2012ApJ...752L...9J}
{Jing}, J., {Park}, S.-H., {Liu}, C., {et~al.} 2012, \apjl, 752, L9

\bibitem[{{Jing} {et~al.}(2015){Jing}, {Xu}, {Lee}, {Nitta}, {Liu}, {Park},
  {Wiegelmann}, \& {Wang}}]{2015RAA....15.1537J}
{Jing}, J., {Xu}, Y., {Lee}, J., {et~al.} 2015, Research in Astronomy and
  Astrophysics, 15, 1537

\bibitem[{{Leka} {et~al.}(2009){Leka}, {Barnes}, {Crouch}, {Metcalf}, {Gary},
  {Jing}, \& {Liu}}]{2009SoPh..260...83L}
{Leka}, K.~D., {Barnes}, G., {Crouch}, A.~D., {et~al.} 2009, \solphys, 260, 83

\bibitem[{{Metcalf}(1994)}]{1994SoPh..155..235M}
{Metcalf}, T.~R. 1994, \solphys, 155, 235

\bibitem[{{Moraitis} {et~al.}(2014){Moraitis}, {Tziotziou}, {Georgoulis}, \&
  {Archontis}}]{2014SoPh..289.4453M}
{Moraitis}, K., {Tziotziou}, K., {Georgoulis}, M.~K., \& {Archontis}, V. 2014,
  \solphys, 289, 4453

\bibitem[{{Pariat} {et~al.}(2017){Pariat}, {Leake}, {Valori}, {Linton},
  {Zuccarello}, \& {Dalmasse}}]{2017A&A...601A.125P}
{Pariat}, E., {Leake}, J.~E., {Valori}, G., {et~al.} 2017, \aap, 601, A125

\bibitem[{{Pesnell} {et~al.}(2012){Pesnell}, {Thompson}, \&
  {Chamberlin}}]{2012SoPh..275....3P}
{Pesnell}, W.~D., {Thompson}, B.~J., \& {Chamberlin}, P.~C. 2012, \solphys,
  275, 3

\bibitem[{{Prior} \& {Yeates}(2014)}]{2014ApJ...787..100P}
{Prior}, C., \& {Yeates}, A.~R. 2014, \apj, 787, 100

\bibitem[{{Schou} {et~al.}(2012){Schou}, {Scherrer}, {Bush}, {Wachter},
  {Couvidat}, {Rabello-Soares}, {Bogart}, {Hoeksema}, {Liu}, {Duvall}, {Akin},
  {Allard}, {Miles}, {Rairden}, {Shine}, {Tarbell}, {Title}, {Wolfson},
  {Elmore}, {Norton}, \& {Tomczyk}}]{2012SoPh..275..229S}
{Schou}, J., {Scherrer}, P.~H., {Bush}, R.~I., {et~al.} 2012, \solphys, 275,
  229

\bibitem[{{Schrijver} {et~al.}(2006){Schrijver}, {De Rosa}, {Metcalf}, {Liu},
  {McTiernan}, {R{\'e}gnier}, {Valori}, {Wheatland}, \&
  {Wiegelmann}}]{2006SoPh..235..161S}
{Schrijver}, C.~J., {De Rosa}, M.~L., {Metcalf}, T.~R., {et~al.} 2006,
  \solphys, 235, 161

\bibitem[{{Sun} {et~al.}(2012){Sun}, {Hoeksema}, {Liu}, {Wiegelmann},
  {Hayashi}, {Chen}, \& {Thalmann}}]{2012ApJ...748...77S}
{Sun}, X., {Hoeksema}, J.~T., {Liu}, Y., {et~al.} 2012, \apj, 748, 77

\bibitem[{{Thalmann} {et~al.}(2011){Thalmann}, {Inhester}, \&
  {Wiegelmann}}]{2011SoPh..272..243T}
{Thalmann}, J.~K., {Inhester}, B., \& {Wiegelmann}, T. 2011, \solphys, 272, 243

\bibitem[{{Valori} {et~al.}(2012){Valori}, {D{\'e}moulin}, \&
  {Pariat}}]{2012SoPh..278..347V}
{Valori}, G., {D{\'e}moulin}, P., \& {Pariat}, E. 2012, \solphys, 278, 347

\bibitem[{{Valori} {et~al.}(2013){Valori}, {D{\'e}moulin}, {Pariat}, \&
  {Masson}}]{2013A&A...553A..38V}
{Valori}, G., {D{\'e}moulin}, P., {Pariat}, E., \& {Masson}, S. 2013, \aap,
  553, A38

\bibitem[{{Valori} {et~al.}(2016){Valori}, {Pariat}, {Anfinogentov}, {Chen},
  {Georgoulis}, {Guo}, {Liu}, {Moraitis}, {Thalmann}, \&
  {Yang}}]{2016SSRv..201..147V}
{Valori}, G., {Pariat}, E., {Anfinogentov}, S., {et~al.} 2016, \ssr, 201, 147

\bibitem[{{Wheatland} {et~al.}(2000){Wheatland}, {Sturrock}, \&
  {Roumeliotis}}]{2000ApJ...540.1150W}
{Wheatland}, M.~S., {Sturrock}, P.~A., \& {Roumeliotis}, G. 2000, \apj, 540,
  1150

\bibitem[{{Wiegelmann}(2004)}]{2004SoPh..219...87W}
{Wiegelmann}, T. 2004, \solphys, 219, 87

\bibitem[{{Wiegelmann} \& {Inhester}(2010)}]{2010A&A...516A.107W}
{Wiegelmann}, T., \& {Inhester}, B. 2010, \aap, 516, A107

\bibitem[{{Wiegelmann} {et~al.}(2006){Wiegelmann}, {Inhester}, \&
  {Sakurai}}]{2006SoPh..233..215W}
{Wiegelmann}, T., {Inhester}, B., \& {Sakurai}, T. 2006, \solphys, 233, 215

\end{thebibliography}
\end{document}